\documentclass{article}

\usepackage{amsmath}
\usepackage{bm}

\usepackage{arxiv}
\usepackage{xcolor}
\usepackage[utf8]{inputenc}
\usepackage[T1]{fontenc}
\usepackage{url}
\usepackage{booktabs}
\usepackage{amsfonts}
\usepackage{nicefrac}
\usepackage{lipsum}
\usepackage{graphicx}
\usepackage{placeins}
\usepackage[authoryear,round]{natbib}
\usepackage{hyperref}
\usepackage{doi}

\title{Mixture of Polyconvex Neural Potentials for Parametric Hyperelasticity: Towards Foundation Material Models}

\author{ \href{https://orcid.org/0000-0002-8282-856X}{\includegraphics[scale=0.06]{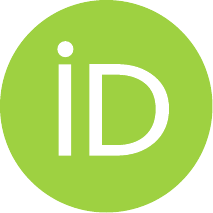}\hspace{1mm}Steven J. Yang} \\
	Cornell University \&  Neural Solid \\
	\texttt{sjy32@cornell.edu} \\
    \And
    \href{https://orcid.org/0009-0006-2956-1989}{\includegraphics[scale=0.06]{orcid.pdf}\hspace{1mm}Govinda Anantha Padmanabha} \\
	Department\\
	École Polytechnique Fédérale de Lausanne (EPFL) \\ \& Cornell University\\
	\texttt{govinda.ananathapadmanabha@epfl.ch} \\
    \And
    \href{https://orcid.org/0000-0002-4579-2676}
    {\includegraphics[scale=0.06]{orcid.pdf}\hspace{1mm}D. Thomas Seidl} \\
    Sandia National Laboratories \\
    \texttt{dtseidl@sandia.gov} \\
    \And
	\href{https://orcid.org/0000-0002-3349-5914}{\includegraphics[scale=0.06]{orcid.pdf}\hspace{1mm}Nikolaos Bouklas} \\
	Cornell University \& Neural Solid \\
	\texttt{nb589@cornell.edu} \\
}
\date{}

\renewcommand{\shorttitle}{Mixture of Polyconvex Neural Potentials for Parametric Hyperelasticity}

\hypersetup{
pdftitle={A template for the arxiv style},
pdfsubject={q-bio.NC, q-bio.QM},
pdfauthor={David S.~Hippocampus, Elias D.~Striatum},
pdfkeywords={First keyword, Second keyword, More},
}

\begin{document}
\maketitle

\begin{abstract}

Hyperelastic constitutive models enable modeling large deformations in elastic solids. In common practice, a strain energy density function is prescribed in advance and model-specific parameters are calibrated from experiments. However, many applications require constitutive models for a family of related materials whose mechanical behavior varies with composition. A fixed constitutive model-form may not capture the full range of behavior across the family, while fitting separate forms does not provide a direct way to predict the response of new compositions. Recent work has developed data-driven constitutive models that learn flexible strain energy functions while incorporating key physical constraints. In this work, we propose using mixtures of convex neural potentials based on input convex neural networks as a modular and data efficient approach to modeling material families. Each potential is convex and monotonic with respect to polyconvex strain invariants, while a conditioning network maps material descriptors to mixture weights. We compare the approach with a monolithic partially input-convex neural network using experimental data from PolyJet 3D-printed materials and a synthetic Gent-type benchmark. Across both benchmarks, the mixture architecture generalized better to material descriptors not seen during training. In the PolyJet experimental benchmark, we showed that the mixture architecture is less sensitive to model hyperparameters, while in the Gent-type benchmark it generalized more reliably with sparse data in the material-descriptor space. These results suggest that representing a material family through a small set of shared convex neural potentials provides a useful structural prior for learning descriptor-dependent constitutive behavior from limited data. 

\end{abstract}

\newpage
\section{Introduction}
\vspace{-1em}
Hyperelastic constitutive models are essential for simulating the mechanical responses of elastic solids undergoing large recoverable deformations. In these models, the constitutive response is specified by a strain energy density function, whose derivatives determine the corresponding stresses \cite{holzapfel2000nonlinear}. More broadly, strain  energy density functions serve as a core component for many inelastic constitutive models, where they characterize the elastically recoverable part of the response while additional evolution laws describe dissipative mechanisms \cite{gurtin_fried_anand_2010}. In many applications, however, the goal is not to model one material at a time, but a family of related materials whose responses vary with composition, microstructure, or other design parameters. Examples include polymer networks with tunable bonds, filled rubbers, segmented copolymers, and multi-material additive manufacturing (\cite{khare2021transition,vidavsky2020tuning,heinrich2002reinforcement,xian2024filled,kang2024efficient,yang2024elucidating,daneshdoost2024structure,slesarenko2018towards}). A common approach is to calibrate analytical constitutive models separately for each material, often selecting the functional form and parameters on a material-by-material basis \cite{ricker2023systematic}. However, using a single functional form requires that it remain appropriate across the material family, while fitting different forms does not provide a predictive model of the family. 

Data-driven constitutive modeling has emerged as an approach for representing material behavior without prescribing fixed analytical constitutive forms in advance \cite{fuhg2025areview}. This is particularly useful for complex material responses, where selecting a particular constitutive form can lead to model-form error. Neural constitutive models have been developed for both elastic and inelastic behavior, including isotropic and anisotropic elasticity, viscoelasticity, and plasticity. These models have been formulated as neural strain energy potentials as well as neural dissipation potentials, neural ODEs, and recurrent neural networks for history-dependent responses (\cite{linden2023neural,linka2021constitutive,linka2023new,flaschel2025convex,jones2026physics,tac2022data,tac2023data,tac2026fully,kalina2026physics,ghavamian2019accelerating}). Critically, physics-augmented neural networks (PANNs) have become a common framework across these approaches, in which requirements such as thermodynamic admissibility, material symmetry, and objectivity are embedded directly into the learned representation~\cite{linden2023neural}. 

For data-driven hyperelastic models, input-convex neural networks (ICNNs) (\citet{amos2017Input}) provide a structured approach for learning strain energy functions~(\citet{klein2022polyconvex, padmanabha2024improving, fuhg2024extreme}). These models often use deformation invariants or related measures as inputs, with stresses derived from the learned strain energy function. Convexity and monotonicity are enforced with respect to selected deformation measures. A common strategy is to choose deformation measures that are individually convex in the deformation gradient $\bm{F}$, $\operatorname{cof}\bm{F}$, and $\det \bm{F}$, such that the learned function is polyconvex in the sense of Ball (\cite{ball1976convexity,hartmann2003polyconvexity}). In ICNNs, convexity is enforced by using convex non-decreasing activation functions and constraining the hidden-to-hidden and output-layer weights to be non-negative, while leaving the input weights unconstrained. Similarly, a Monotonic Neural Network (MNN) enforces monotonicity by using non-decreasing activations and non-negative weights for all layers \cite{klein2501neural}.

To represent families of hyperelastic materials within a single model, studies have employed partial input-convex neural networks (pICNNs) and partial monotonic neural networks conditioned on material descriptors (\cite{klein2023parametrized,yang2025physics,klein2026Neural,jadoon2026multiscale}). These descriptors are incorporated through an unconstrained conditioning network branch that modulates the learned strain energy function. Such models have enabled optimization and inverse design of structures with spatially varying material compositions (\cite{luo2026towards,jadoon2026multiscale}). In typical pICNN implementations, latent features derived from material descriptors are injected throughout the convex branch. The result is a single constrained architecture that must capture a range of material behaviors within a shared parameterization. At the same time, the non-negative weight constraints used to enforce convexity in ICNNs can introduce additional optimization difficulties~\citet{amos2017Input}.
 
While existing PANN constitutive models are typically formulated as monolithic architectures, additive and modular neural networks have been shown to improve interpretability, reduce optimization interference, and facilitate the reuse of learned components~\cite{pfeiffer2023modular}. In mixture-of-experts (MoE) architectures, predictive mappings are represented through a collection of specialized subfunctions whose contributions are determined by a routing mechanism. Depending on the routing strategy, a subset of subfunctions may be selected through hard gating or combined through soft routing weights. MoEs have been successfully applied to a variety of machine learning fields including computer vision~\cite{riquelme2021scaling}, natural language processing~\cite{shazeer2017outrageously}, and large language models~\cite{zhang2025mixture}. More recently, scientific machine learning approaches, such as operator learning frameworks~(\cite{sharma2024ensemble,deighan2025mixture}), have begun to incorporate MoE-based architectures. \citet{agarwal2021neural} introduced Neural Additive Models (NAMs), which provide predictions as a sum of input feature-specific subnetworks. This decomposition allows model components to be independently analyzed while maintaining competitive performance. More broadly, such architectures suggest that predictive mappings can be constructed from additive combinations of shared learned functions. This perspective is particularly appealing for constitutive modeling, where related materials may share common constitutive components while differing in their overall mechanical response. 

Here, we propose and systematically investigate the use of mixtures of convex neural potentials for modeling families of hyperelastic materials. Instead of using one monolithic pICNN, the model learns several shared strain energy components whose weighted sum defines the final potential. Each strain energy component uses only polyconvex strain invariant measures as inputs and satisfies convexity and monotonicity constraints. A separate conditioning network maps the material descriptors to non-negative mixture weights. Critically, no particular functional form are imposed on how the mixture weights depend on material descriptors, allowing the model to learn this relationship flexibly from data. Because the final potential is formed as a non-negative linear combination of convex-strain invariant components, this architecture preserves polyconvexity. This construction is motivated by the idea that varying material behaviors can be represented through a relatively small set of shared strain energy components. At the same time, this modular decomposition may reduce optimization interference and provide a more interpretable representation. 

Related constructions have recently been explored in constitutive learning. Constitutive artificial neural networks (CANNs) have been combined with feature networks to weight individual constitutive terms based on water content in hydrated soft materials \cite{boes2026mechanics}. Similarly, DeepONet-based neural operators have been developed to combine latent strain energy features through a weighted sum to infer hyperelastic potentials from displacement-field and reaction-force measurements \cite{flaschel2026neural}. The present work focuses on systematically investigating whether a low-dimensional mixture of shared convex strain energy components improves generalization across material descriptors compared with monolithic pICNN conditioning.

In this paper, Section \ref{sec:hyperelasticity} introduces the constitutive theory for hyperelasticity and the polyconvex invariant measures used to define the strain energy function. Section \ref{sec:constrained_NN} reviews the constrained neural network architectures used in this work, including ICNNs, MNNs, and the monolithic pICNN baseline. Section \ref{sec:mixture_network} presents the proposed mixture architecture. Section \ref{sec:model_evaluation} evaluates the proposed approach on an experimental multi-material PolyJet dataset and a synthetic Gent-type benchmark on unseen material descriptors and deformation modes. Finally, Section \ref{sec:conclusion} summarizes the main findings and discusses limitations and directions for future work. 



\section{Constitutive Theory for Hyperelasticity}
\label{sec:hyperelasticity}

A hyperelastic material is defined by its strain energy density function $\psi(\bm{F})$, where $\bm{F} = \partial\bm{x}/\partial\bm{X}$ is the deformation gradient mapping a material point $\bm{X}$ in the reference configuration to its deformed position $\bm{x}(\bm{X})$, with $\det \bm{F} > 0$. For isotropic hyperelasticity, objectivity implies that the strain energy depends only on the invariants of the right Cauchy-Green tensor $\bm{C} = \bm{F}^{T}\bm{F}$. The invariants are
\begin{equation}
I_{1} = \mathrm{tr}(\bm{C}), \qquad
I_{2} = \frac{1}{2}\Big[ \mathrm{tr}(\bm{C})^{2} - \mathrm{tr}(\bm{C}^{2}) \Big], \qquad 
I_{3} = \det(\bm{C}) = J^{2}, \qquad 
\end{equation}
where $J = \det{\bm{F}}$ denotes the volumetric change. Accordingly, the strain energy may be written in the form $\psi(I_1,I_2,I_3)$, or equivalently $\psi(I_1,I_2,J)$. 

We focus on incompressible and nearly incompressible materials and adopt the standard isochoric--volumetric split of the strain energy,
\begin{equation}
\psi = \psi_{\mathrm{iso}}(\bar{I}_{1},\bar{I}_{2}) + \psi_{\mathrm{vol}}(J), 
\end{equation}
where 
\begin{equation}
\bar{\bm{C}} = J^{-2/3}\bm{C} , \qquad
\bar{I}_1 = \mathrm{tr}{\bar{\bm{C}}} = J^{-2/3}I_{1}  , \qquad
\bar{I}_2 = \frac{1}{2}\Big[ (\mathrm{tr}\bar{\bm{C}})^{2} - \mathrm{tr}(\bar{\bm{C}}^{2}) \Big] = J^{-4/3}I_{2}. \qquad
\end{equation}
For incompressible materials the constraint $J=1$ is enforced by a pressure field acting as a Lagrange multiplier, and the energy reduces to the isochoric contribution. For nearly incompressible materials, volumetric changes are penalized through the volumetric energy $\psi_{\mathrm{vol}}(J)$. In the present work, the neural network learns only the isochoric contribution $\psi_{\mathrm{iso}}$.  

In addition to objectivity and isotropy, we further restrict attention to strain energy functions that are polyconvex, as defined by Ball. This choice promotes a stable and physically reasonable elastic response. Polyconvexity means that the strain energy can be expressed as a convex function of $\bm{F}$, $\operatorname{cof}\bm{F}$, and $\det\bm{F}$. In practice, this is achieved by constructing the strain energy as a convex, non-decreasing function of polyconvex invariant measures. The isochoric invariant $\bar{I}_{1}$ is polyconvex, whereas $\bar{I}_{2}$ is not polyconvex. However, the quantity $\bar{I}_{2}^{3/2}$ is polyconvex (\cite{hartmann2003polyconvexity,klein2026Neural}).

Since $\bar{I}_{1} \geq 3$ and $\bar{I}_{2} \geq 3$, the shifted quantities $\xi_1 = \bar{I}_{1} - 3$ and $\xi_2 = \bar{I}_{2}^{3/2} - 3\sqrt{3}$ are non-negative. Polyconvexity is preserved under the addition of constants. We therefore construct the isochoric strain energy in the form
\begin{equation}
\psi_{\mathrm{iso}}(\xi_1,\xi_2),
\end{equation}
where $\psi_{\mathrm{iso}}$ is convex and non-decreasing in each argument. This ensures that the isochoric energy attains its minimum at the reference configuration and, more generally, for purely volumetric deformations. 

The second Piola–Kirchhoff stress from the isochoric strain energy density function can be obtained by
\begin{equation}
\label{eq:PK2_stress}
\bm{S} = 2\frac{\partial \psi}{\partial \bm{C}}.
\end{equation}
With the split $\psi = \psi_{\mathrm{iso}}(\xi_1,\xi_2) + \psi_{\mathrm{vol}}(J)$, this gives $\bm{S} = \bm{S}_{\mathrm{iso}} + \bm{S}_{\mathrm{vol}}$. The isochoric stress is given by
\begin{equation}
\label{eq:PK2_stress_iso}
\bm{S}_{\mathrm{iso}} = 2 \bigg(\frac{\partial \psi_{\mathrm{iso}}}{\partial \xi_{1}}\frac{\partial \xi_{1}}{\partial \bm{C}} + \frac{\partial \psi_{\mathrm{iso}}}{\partial  \xi_{2}}\frac{\partial \xi_{2}}{\partial \bm{C}}\bigg) 
\end{equation}
where
\begin{equation}
\label{eq:xi_derivatives}
\frac{\partial \xi_{1}}{\partial \bm{C}} = J^{-2/3}\Big(\bm{1} -\frac{1}{3}I_{1}\bm{C}^{-1} \Big) ,  \qquad
\frac{\partial \xi_{2}}{\partial \bm{C}} =
J^{-2} \bigg[ \frac{3}{2} I_{2}^{1/2} \big(I_{1}\bm{1} - \bm{C}\big) - I_{2}^{3/2}\bm{C}^{-1}\bigg].\qquad
\end{equation}
At the reference configuration, $\bm{C}=\bm{1}$, both derivatives vanish, ensuring that the isochoric stress is zero without requiring an additional stress-free constraint. The corresponding Cauchy stress tensor is obtained from the second Piola--Kirchhoff stress through
\begin{equation}
\label{eq:PK2_to_Cauchy}
\bm{\sigma} = J^{-1} \bm{F} \bm{S} \bm{F}^{T}.
\end{equation}

All stresses reported throughout this work are Cauchy stresses.
In the model evaluations below, material responses are modeled as incompressible, so that $J=1$, and the neural network learns only the isochoric strain energy. The isochoric stress response is computed, and the in-plane Cauchy stress comlonents are reported after enforcing through-thickness plane-stress condition $\sigma_{33} = 0$.

To represent families of hyperelastic materials within a single constitutive model, we extend this representation to depend on a vector of material descriptors $\bm{v}$, yielding the parametric form
\begin{equation}
\psi_{\mathrm{\mathrm{iso}}}(\xi_{1},\xi_{2},\bm{v}).
\end{equation}
In this work, the isochoric strain energy $\psi_{\mathrm{iso}}$ is represented using neural network parameterizations that enforce convexity and monotonicity with respect to ($\xi_{1}$,$\xi_{2}$), while remaining unconstrained in $\bm{v}$.   

\section{Constrained Neural Network Architectures}
\label{sec:constrained_NN}

\subsection{Convex and Monotonic Neural Networks}
\label{sec:convex_monotonic_NN}

Consider a feed-forward neural network that maps an input vector
$\bm{x}_{0}$ to an output $\bm{y}$ through a sequence of hidden states
$\{\bm{x}_\ell\}_{\ell=1}^{L-1}$. 
The forward propagation is defined recursively as:
\begin{equation}
\begin{aligned}
\bm{x}_{1} &= \sigma_{1}\!\left(\bm{W}_{1} \bm{x}_{0} + \bm{b}_{1}\right),
\\[4pt]
\bm{x}_{\ell} &= \sigma_{\ell}\!\left(
\bm{W}_{\ell} \bm{x}_{\ell-1} 
+ \widetilde{\bm{W}}_{\ell} \bm{x}_{0}\,
+ \bm{b}_{\ell}
\right),
\qquad \ell = 2,\ldots,L-1,
\\[4pt]
\bm{y} &= \bm{W}_{L} \bm{x}_{L-1}
+ \widetilde{\bm{W}}_{L} \bm{x}_{0}\,
\end{aligned}
\label{eq:forward}
\end{equation}

Here, $\bm{W}_\ell$ and $\widetilde{\bm{W}}_\ell$ denote weight matrices, $\bm{b}_\ell$ are bias vectors, and $\sigma_\ell(\cdot)$ represent element-wise activation functions. All weights and biases collectively define the trainable parameter set $\bm{\theta} = \{\{ \bm{W}_\ell \}_{\ell=1}^{L} \cup \{\widetilde{\bm{W}}_\ell \}_{\ell=2}^{L} \cup \{\bm{b}_\ell\}_{\ell=1}^{L-1}\}$.

An input convex neural network (ICNN) is a neural architecture designed so that the network output $\bm{y}$ is convex with respect to the input $\bm{x}_0$. Using the general architecture defined in Eq.~\eqref{eq:forward}, convexity is enforced through constraints on the activation functions and certain network parameters. In particular, the activation functions $\sigma_\ell(\cdot)$ (for $\ell=1,\dots,L-1$) are required to be convex and non-decreasing (e.g., \emph{Softplus}, \emph{ReLU}). Moreover, the hidden-state and output-layer weight matrices $\bm{W}_\ell$ (for $\ell=2,\dots,L$) are constrained to be element-wise non-negative. Together, these conditions preserve convexity through composition with convex, non-decreasing activations and non-negative linear combinations of those activations. For ICNNs, the first-layer weights $\bm{W}_{1}$, input pass-through weights $\widetilde{\bm{W}}_{\ell}$, and biases $\bm{b}_\ell$ are unconstrained. 

A monotonic neural network (MNN) guarantees that the network output $\bm{y}$ is monotonic with respect to the input $\bm{x}_0$. For the architecture in Eq.~\eqref{eq:forward}, a non-decreasing mapping is achieved by requiring the activation functions $\sigma_\ell(\cdot)$ to be non-decreasing. The weight matrices $\bm{W}_\ell$ (including the first-layer, $\ell=1$) and $\widetilde{\bm{W}}_\ell$ are constrained to be element-wise non-negative. The biases $\bm{b}_\ell$ remain unconstrained.

For the polyconvex neural potentials considered in this work, since we use shifted invariants, the network must satisfy both input convexity and input monotonicity. Accordingly, the activation functions $\sigma_\ell(\cdot)$ are required to be convex and non-decreasing, the weight matrices $\bm{W}_\ell$ and $\widetilde{\bm{W}}_\ell$ are constrained to be element-wise non-negative, and no constraints are imposed on the biases $\bm{b}_\ell$.

\subsection{Monolithic Partial-Input Convex Neural Network Potential (Baseline)}
\label{sec:monolithic_pICNN}

Several recent works extend ICNNs to parametric neural potentials by introducing additional conditioning inputs not subject to convexity and monotonicity constraints. This is commonly implemented using the partial-input convex neural network (pICNN) architecture introduced by \citet{amos2017Input}, which was subsequently adopted for constitutive modeling applications (\citet{klein2023parametrized, yang2025physics, jadoon2026input}). We adopt this approach as the baseline architecture. 
Let the input vector be partitioned as $\bm{x}_{0} =[\bm{u},\bm{v}]$, where $\bm{u}$ contains  polyconvex strain invariant measures and $\bm{v}$ contains material descriptor (conditioning) variables. These conditioning variables are processed through a separate unconstrained neural pathway whose hidden features are injected into the convex network. In this work, the conditioning network is implemented as a feed-forward neural network producing hidden features $\bm{z}_{\ell}(\bm{v})$ defined by:
\begin{equation}
\begin{aligned}
\bm{z}_{1} &= \phi_{1}\!\left(\bm{V}_{1} \bm{v} + \bm{c}_{1}\right), \\
\bm{z}_{\ell} &= \phi_{\ell}\!\left(\bm{V}_{\ell} \bm{z}_{\ell-1} + \bm{c}_{\ell}\right),
\qquad \ell = 2,\dots,L-1.
\end{aligned}
\end{equation}
$\bm{V}_{\ell}$, $\bm{c}_{\ell}$, and $\phi_\ell(\cdot)$ denote the weights, biases, and activation functions of the conditioning network. These hidden features are injected into the monotonic convex network as follows: 
\begin{equation}
\begin{aligned}
\bm{x}_{1} &= \sigma_{1}\!\left(\bm{W}_{1} \bm{u} + {\widetilde{\bm{V}}}_{1} \bm{v} + \bm{b}_{1}\right), \\
\bm{x}_{\ell} &= \sigma_{\ell}\!\left(
\bm{W}_{\ell} \bm{x}_{\ell-1}
+ \widetilde{\bm{W}}_{\ell} \bm{u}\,
+ \widetilde{\bm{V}}_{\ell} \bm{z}_{\ell-1}
+ \bm{b}_{\ell}
\right),
\qquad \ell = 2,\dots,L-1, \\
\bm{y} &=
\bm{W}_{L} \bm{x}_{L-1}
+ \widetilde{\bm{W}}_{L} \bm{u}.
\end{aligned}
\end{equation}

Here, ${\bm{W}}_\ell$ and $\widetilde{\bm{W}}_\ell$ denote the weights and input pass-through weights of the monotonic convex network, which are constrained to be non-negative. The activation functions $\sigma_\ell(\cdot)$ are constrained to be convex and non-decreasing. $\widetilde{\bm{V}}_\ell$ represents the weights that connect the conditioning network to the monotonic convex network. ${\bm{b}}_\ell$ are the biases. In the general pICNN formulation, $\bm{y}$ denotes the vector-valued network output. Because the conditioning features $\bm{z}_{\ell-1}$ are injected at every layer of the convex network, each hidden state $\bm{x}_{\ell}$ depends jointly on both inputs $\bm{u}$ and $\bm{v}$. The conditioning variables therefore influence the convex representation throughout the depth of the network. We refer to this formulation as a monolithic pICNN potential.

In this work, the monolithic pICNN defines a scalar-valued function $\hat{y}(\bm{u},\bm{v})$ representing the strain energy response. The isochoric strain energy is defined as
\begin{equation}
\psi_{\mathrm{iso}} =
\hat{y}(\bm{u},\bm{v}) - \hat{y}(\bm{u}_0,\bm{v}),
\end{equation}
where $\bm{u}$ denotes the vector of isochoric invariant-based measures and $\bm{u}_0$ denotes the same measures evaluated in the undeformed material configuration.

\section{Mixture of Convex Potentials}
\label{sec:mixture_network}
In the monolithic pICNN architecture described in Section~\ref{sec:monolithic_pICNN}, the strain energy is represented by a neural network whose internal layers receive conditioning features. As a result, the convex representation must simultaneously accommodate both strain invariants and material descriptors within a single constrained network. 

\begin{figure}[!h]
    \centering
    \includegraphics[width=0.8\linewidth]{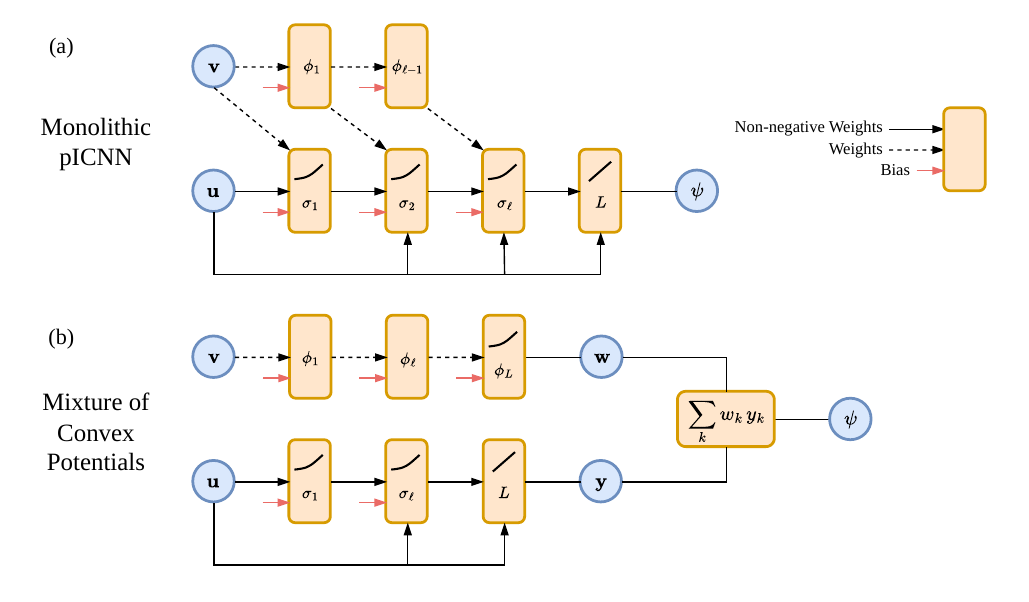}
    \caption{Schematics of (a) monolithic partial input convex neural network (pICNN) and (b) proposed mixture of convex potentials architecture. For visual clarity, panel (b) writes the mixture using the outputs $y_k$; the implemented formulation uses $\hat{y}_k(\bm{u}) - \hat{y}_k(\bm{u}_0)$ where $\bm{u}_0$ is the invariant-based measures in the undeformed configuration.}
    \label{fig:framework}
\end{figure}

We compare this monolithic architecture with a mixture of convex neural potentials. In the mixture architecture, the model produces $K$ convex components $\hat{y}_{k}(\bm{u})$, each depending only on the isochoric invariant-based measures $\bm{u}$. These components are produced by a neural network satisfying the convexity and monotonicity constraints described in Section~\ref{sec:convex_monotonic_NN}. The material descriptor variables $\bm{v}$ are processed by a separate neural network that produces non-negative mixture weights $\hat{w}_{k}(\bm{v})$. The resulting strain energy is expressed as:

\begin{equation}
\label{eq:mixture_energy}
\psi_{\mathrm{iso}} = \sum_{k=1}^{K} \hat{w}_k(\bm{v})
\left[
\hat{y}_{k}(\bm{u})- \hat{y}_{k}(\bm{u}_0)
\right].
\end{equation}
Here, $\bm{u}_0$ denotes the invariant-based measures evaluated in the undeformed material configuration. Because each component $\hat{y}_k(\bm{u})$ is convex with respect to $\bm{u}$ and the mixture weights satisfy $\hat{w}_k(\bm{v}) \ge 0$, the resulting strain energy is a non-negative linear combination of convex functions and therefore remains convex. Thus, the formulation preserves polyconvexity because $\bm{u}$ is constructed from polyconvex invariant measures. At the reference configuration the derivatives of the isochoric invariant measures vanish. Since the strain energy is a weighted sum of these isochoric energy components, the isochoric stress remains zero regardless of the material-dependent mixture weights. Figure~\ref{fig:framework} illustrates the monolithic pICNN architecture and the proposed mixture architecture.

\FloatBarrier
\section{Model Evaluation}\label{sec:model_evaluation}

\subsection{Network Architecture, Training, and Evaluation Details}
\label{sec:general_network_training_evaluation}

All neural network models are implemented in PyTorch using fully connected feed-forward layers. These models take deformation invariant measures and material descriptors as inputs and define a scalar strain energy potential. Predicted stresses are then obtained by automatic differentiation \cite{paszke2017automatic} of the learned strain energy potential with respect to the invariant measures. The \emph{Softplus} activation function~\citep{dugas2009incorporating} is used in the hidden layers of both monolithic pICNN and mixture ICNN models. In the convex branches, its smooth, convex, and non-decreasing form preserves the convexity and monotonicity constraints and yields smooth stress responses. The same activation is used in the conditioning branches for architectural consistency. 

Training is performed by minimizing the mean-squared error (MSE) loss between predicted and ground-truth stress components. Model parameters are optimized using AdamW (\cite{loshchilov2017decoupled}), with regularization applied through its decoupled weight decay term. The same optimizer configuration was used for all models, with learning rate  $10^{-3}$ and weight decay coefficient $10^{-4}$. The AdamW parameters $\beta_{1}$, $\beta_{2}$, $\epsilon$ were set to their default values of 0.9, 0.999, and $10^{-8}$, respectively. 

Model accuracy is evaluated using the root-mean-squared error (RMSE) and a test $R^2$ prediction score as defined below. For some loading modes, certain stress components remain zero over all evaluated samples. These components are excluded from the error metrics. Let $\bm{s}_{n}^{\mathrm{true}}$ and $\bm{s}_{n}^{\mathrm{pred}} \in \mathbb{R}^{M}$ denote the active stress components for evaluated sample $n$, where $M$ is the number of active components. For $N$ evaluated samples, the RMSE is defined as
\begin{equation}
\mathrm{RMSE} = \Bigg( \frac{1}{NM} \sum_{n=1}^{N} \left\| \bm{s}_{n}^{\mathrm{pred}} - \bm{s}_{n}^{\mathrm{true}} \right\|_{2}^{2} \Bigg)^{1/2}.
\end{equation}
The test $R^2$ score is computed separately for each active stress component and then uniformly averaged:
\begin{equation}
\text{Test }R^{2} \text{ Score} = \frac{1}{M} \sum_{m=1}^{M} \Bigg(1 - \frac{\mathrm{SSE}_{m}}{\mathrm{SST}_{m}} \Bigg), \quad \mathrm{SSE}_{m} = \sum_{n=1}^{N} (s_{n,m}^{\mathrm{pred}} -  s_{n,m}^{\mathrm{true}} )^{2}, \quad \mathrm{SST}_{m} = \sum_{n=1}^{N} (s_{n,m}^{\mathrm{true}} - \bar{s}_{m}^{\mathrm{true}})^{2}
\end{equation}
where $\bar{s}_{m}^{\mathrm{true}}$ is the mean ground truth value of the active stress component. This $R^2$-type score is used as a normalized prediction-error on the test sets.

\subsection{Experimental Multi-material 3D Printing Benchmark}
\label{sec:experimental_multimaterial_3d_printing}

\textbf{Dataset and Task:} We consider a multi-material Stratasys PolyJet 3D printing dataset consisting of six digital materials (DMs): A, DM40, DM50, DM60, DM70, and DM85 \cite{yang2025physics}. In PolyJet printing, DMs are produced by spatially mixing droplets of photopolymer resins during printing. In the present dataset, the proportions of Agilus and Digital ABS are varied to obtain a family of related materials with distinct mechanical responses. Agilus is a soft elastomeric resin, while Digital ABS is a stiffer resin system composed of Stratasys RGD515 and RGD531. The material labels follow the original dataset, where A denotes the nominal Agilus material, while the numerical suffix after DM indicates a nominal Shore A hardness value. Similar DM datasets have been previously used to develop descriptor-conditioned constitutive models based on neural networks (\cite{klein2026Neural,yang2025physics}). The present work considers only uniaxial tension acquired at a single strain rate. The polymer mixing ratio provides a physically meaningful material descriptor, making this dataset a useful benchmark for assessing generalization to unseen material compositions. 

For this dataset, we learn a descriptor-conditioned strain energy function of the form $\psi_{\mathrm{iso}}(\xi_{1},c)$, where $\xi_{1} = \bar{I}_{1} - 3$ and $c$ is a scalar material composition descriptor. The deformation dependence is modeled using only the first-invariant-based measure because this benchmark considers only uniaxial tension data. This avoids dependence on multiple deformation invariants, whose independent contributions to the strain energy cannot be determined from uniaxial tension data alone. The descriptor $c$, listed in Table~\ref{tab:polyjet_materials}, is the normalized Digital ABS mass fraction consumed during printing  (rather than direct measurements of the final specimens), with $c=0$ corresponding to composition A and $c=1$ corresponding to DM85. This input scaling is done to improve numerical conditioning during training, and the architecture itself does not restrict the descriptor space to any particular range.

\begin{table}[!h]
\centering
\small
\caption{Material compositions used in the multi-material Stratasys PolyJet 3D printing dataset. Agilus and Digital ABS contents are reported as mass percentages based on resin consumption during printing, and $c$ is the composition descriptor used to condition the model.}
\label{tab:polyjet_materials}
\begin{tabular}{llll}
\toprule
 Material    & \% Agilus & \% Digital ABS & $c$    \\ \toprule
A    & 75.2      & 24.8           & 0.0    \\
DM40 & 74.8      & 25.2           & 0.0508 \\
DM50 & 74.1      & 25.9           & 0.1352 \\
DM60 & 72.8      & 27.2           & 0.2896 \\
DM70 & 70.3      & 29.7           & 0.6050 \\
DM85 & 67.1      & 32.9           & 1.0    \\ \bottomrule
\end{tabular}
\end{table}

We use a leave-one-out procedure to evaluate generalization across material compositions. In each train--test split, five compositions are used for training and the remaining one is held out for testing. The goal is to predict the full uniaxial tensile stress response of the held-out material. This process is repeated for all six compositions across three independent training runs, giving 18 evaluations for each model configuration. Given the limited data, this protocol provides a more comprehensive assessment of composition generalization than any single train--test split.

\textbf{Model and Training Details:} The model comparison includes the monolithic pICNN baseline and the proposed mixture ICNN architecture at two network sizes, denoted small (S) and large (L). The monolithic pICNN injects conditioning features additively into the convex branch, requiring the conditioning and convex hidden layers to match in width and depth. The small monolithic pICNN uses conditioning and convex hidden-layer widths of $[15,15]$ and has 811 trainable parameters, while the large monolithic pICNN uses widths of $[20,20]$ and has 1381 trainable parameters.  

For the mixture ICNN, the strain energy potential is constructed as a non-negative weighted sum of shared convex energy functions. For this benchmark, we use three mixture components to provide a low-dimensional representation of strain energy across material compositions, without assuming that the number of latent components must match the number of physical constituents. The small mixture model uses convex hidden-layer widths of $[20,20]$ and conditioning hidden-layer width $[10]$, giving 596 trainable parameters. The large mixture model uses convex hidden-layer widths of $[30,30]$ and conditioning hidden-layer width $[15]$, giving 1191 trainable parameters.

The \emph{Softplus} is given as $ \text{Softplus}(x) = \frac{1}{\beta} * \log(1 + \exp(\beta * x))$ and the $\beta$ parameter is varied to assess its effect on model performance. Larger values of $\beta$ produce a sharper activation that more closely approximates \emph{ReLU}. The conditioning branches use a fixed value of $\beta=1.0$, while the convex branches use $\beta \in \{0.5,1.0,2.0\}$. The two architectures, two network sizes, and three convex-branch $\beta$ values give 12 model configurations. Each model is trained for 10,000 epochs using 64 samples along the stress-stretch curve for each material, and a mini-batch size of 64. 

\begin{figure}[!h]
    \centering
    \includegraphics[width=0.8\linewidth]{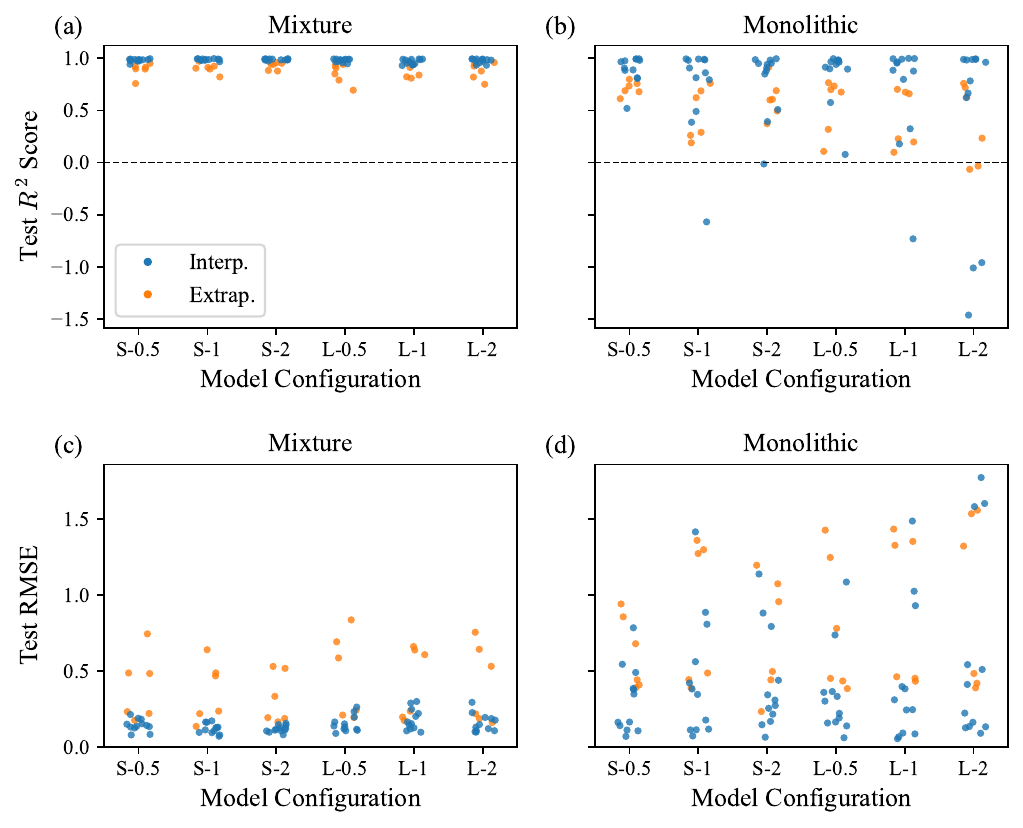}
    \caption{Test performance for the PolyJet 3D printed material dataset, evaluated using (a, b) $R^2$ scores and (c, d) root mean squared errors (RMSE). Results are shown for (a, c) mixture ICNN models and (b, d) monolithic pICNN models. The labels S and L denote the small and large network-size configurations, and the number denotes the convex-branch \emph{Softplus} parameter $\beta$. Each model configuration has $6$ extrapolation and $12$ interpolation test cases.}
    \label{fig:mm3dp_R2_jitter}
\end{figure}

\textbf{Results and Discussion:} The PolyJet 3D printed material dataset provides an experimental test of whether the proposed mixture model can generalize across material compositions. In each leave-one-out split, the models are trained on five compositions and used to predict the response of a held-out material. We compare the mixture ICNN with the monolithic pICNN baseline to assess whether a small set of shared convex energy components provides an effective inductive bias for modeling families of related materials, given a limited amount of data. Figure~\ref{fig:mm3dp_R2_jitter} shows test $R^2$ scores in panels (a, b) and test root mean squared errors (RMSE) in panels (c, d) for held-out material compositions, comparing the mixture ICNN (a, c) and monolithic pICNN (b, d) models. In this work, we test our models with interpolated and extrapolated data that is unseen during training. We refer to held-out DM40--DM70 splits as interpolation because their composition descriptors lie within the range of descriptors seen during training. DM85 and A are treated as extrapolation test data because their descriptors lie outside of the training range. In Figure~\ref{fig:mm3dp_R2_jitter}(a, c), the mixture ICNN produces high test $R^{2}$ scores indicating accurate held-out predictions overall. The interpolation cases exhibit $R^{2}$ scores clustered near one and consistently low RMSE values, while the extrapolation cases show greater variability but still maintain strong predictive performance. In contrast, Figure~\ref{fig:mm3dp_R2_jitter}(b, d) shows that the monolithic pICNN exhibits a much wider spread in both test $R^{2}$ scores and RMSE values. Although the monolithic pICNN achieves high $R^{2}$ scores and low RMSE values for some splits, especially for interpolation, its performance is less reliable overall with most extrapolation cases and some interpolation cases having test $R^{2}$ scores below 0.75. The averaged error metrics reported in Appendix Table~\ref{tab:mm3dp_metrics} provide values consistent with these observations. 

The predictive performance of the two architectures varies differently with network size and the convex-branch \emph{Softplus} $\beta$ parameter. The monolithic pICNN performs best using the smoother $\beta=0.5$ activation, while increasing network size from small to large does not improve overall performance. In contrast, the mixture ICNN shows strong held-out performance across the tested configurations. This suggests that representing the material family through a small set of shared strain energy components imposes a useful constraint, making the architecture less sensitive to particular choices of network size or activation smoothness. Critically, the final train RMSE values are comparable across architectures (see Appendix Table~\ref{tab:mm3dp_metrics}). Thus, the held-out performance gap is not simply a training-fit issue. The loss histories in Appendix Figure~\ref{fig:mm3dp_training_history} show smooth training-loss convergence across all configurations. However, several low-performing monolithic pICNN configurations maintain high test loss throughout training, indicating poor generalization. In contrast, mixture ICNN models show more stable and consistent test-loss behavior.

\begin{figure}[!h]
    \centering
    \includegraphics[width=0.8\linewidth]{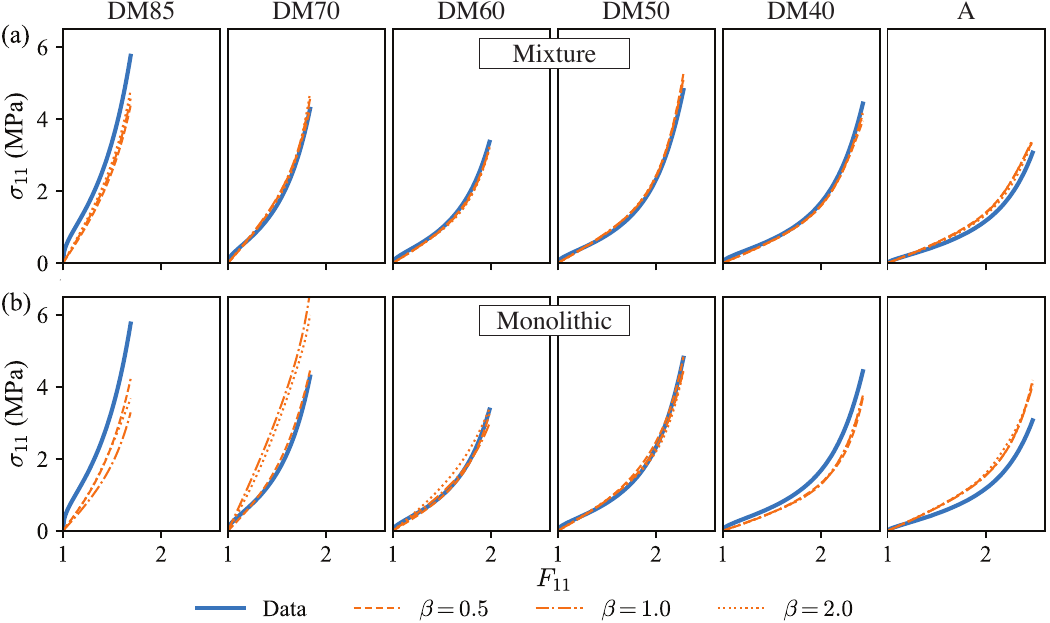}
    \caption{Leave-one-composition-out stress--stretch predictions for the PolyJet dataset. In each panel, the displayed composition is excluded from training, and the model is trained on the other five compositions. Results are shown for the (a) mixture ICNN and (b) monolithic pICNN architectures with the small network-size configuration. Blue curves denote experimental data, and orange curves denote model predictions using convex-branch \emph{Softplus} parameters $\beta \in \{0.5,1.0,2.0\}$.}
    \label{fig:mm3dp_stress_stretch}
\end{figure}

Stress-stretch predictions provide a visual context for interpreting the error metric trends. As a representative example, Figure~\ref{fig:mm3dp_stress_stretch} shows leave-one-out split predictions of $\sigma_{11}$--$F_{11}$ responses for a fixed seed, small network size, and convex-branch \emph{Softplus} parameters $\beta \in \{0.5,1.0,2.0\}$. The mixture ICNN predictions follow the ground-truth data more closely across material compositions. For each split, the entire stress--stretch dataset of the displayed composition is excluded from training, and the model is trained on the other five compositions. The monolithic pICNN results are more sensitive to $\beta$ and show larger deviations in several cases. For example, the DM70 split performs poorly for $\beta=1.0$ and $\beta=2.0$, while $\beta=0.5$ gives a much better prediction. Larger deviations are also observed for A and DM40. For DM85, the ground-truth data includes a small initial bump at low deformation followed by reduced tangent stiffness, and both architectures do not reproduce this feature. This softening behavior is often difficult to capture with convex strain energy functions. Related work has examined other constraints choices for neural hyperelastic models, including monotonicity-only formulations that relax convexity requirements \cite{klein2501neural}. 

\begin{figure}[!h]
    \centering
    \includegraphics[width=0.9\linewidth]
    {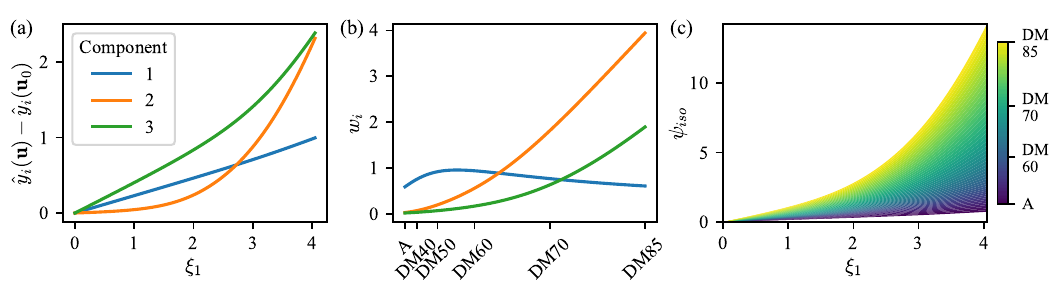}
    \caption{Learned mixture decomposition for a representative PolyJet training run: (a) latent strain energy components, (b) composition-dependent mixture weights, and (c) resulting material-dependent strain energy functions.}
    \label{fig:mm3dp_learned_mixture_energy}
\end{figure}

Figure~\ref{fig:mm3dp_learned_mixture_energy} illustrates one learned mixture decomposition from a representative training run. This example corresponds to a run with DM40 held out, using the small mixture network and convex-branch \emph{Softplus} parameter $\beta=1.0$. This decomposition consists of latent strain energy components and composition-dependent mixture weights, which together define the final material-dependent strain energy function. The learned components and weights should not be interpreted as uniquely identified physical quantities. Different seeds, data splits, or hyperparameter choices could lead to different latent representations. Nonetheless, it provides a useful post hoc view of how the mixture model organizes the learned response for the material family. 

In Figure~\ref{fig:mm3dp_learned_mixture_energy}(a), the latent strain energy components are plotted as functions of the invariant measure $\xi_{1} = \bar{I}_{1} - 3$. In this example, the learned components specialize into distinct energy shapes. These include an approximately linear energy in $\xi_{1}$, corresponding to a neo-Hookean-like response (component 1 shown in blue), and a finite-extensibility-like response with a rapid upturn at larger deformation (component 2 shown in orange). Component 3 (shown in green) shows a mild nonlinear upturn. The corresponding mixture weights in Figure~\ref{fig:mm3dp_learned_mixture_energy}(b) show how these latent energy shapes are combined. Because the non-negative mixture weights are not normalized, they affect both the relative contribution and the overall scale of each latent component. Component 1 remains active and relatively constant across the full composition range, whereas the mixture weights on components 2 and 3 increase from A to DM85. This indicates that the model retains a shared neo-Hookean-like contribution across the material family while increasing the finite-extensibility contribution as the Digital ABS content increases. Figure~\ref{fig:mm3dp_learned_mixture_energy}(c) shows the resulting material-dependent strain energy function. The learned energy varies smoothly across material composition and reflects changes in both the relative scale and curvature with respect to $\xi_{1}$. Here, the specialization of latent components is not promoted by dedicated loss terms or constraints, but emerges from fitting the mixture architecture to the available data. Together with the improved held-out generalization shown above, these results support the idea that related materials can be modeled effectively using a small set of shared latent strain energy components. 

\subsection{Synthetic Gent-Type Benchmark}

\textbf{Benchmark Definition:} To complement the experimental dataset, we construct a synthetic benchmark based on a modified Gent strain energy function. Unlike the experimental dataset, which contains only uniaxial tensile data, the synthetic dataset includes multiple deformation modes and therefore tests learning of a richer isotropic constitutive response. We generate a synthetic material family by prescribing mappings from material descriptors to constitutive model parameters. The mapping is chosen to create regions of rapidly varying constitutive behavior, providing a challenging setting for descriptor-conditioned constitutive learning. We use this synthetic dataset to investigate the effects of training-set size and to evaluate generalization across both material descriptors and loading modes.

The modified Gent strain energy is defined as 
\begin{equation}
    \psi_{\mathrm{iso}} = -C_{1} J_{m} \ln \Big(1 - \frac{\bar{I}_1 - 3}{J_{m}}  \Big) + C_{2} \big(\bar{I}_{2}^{3/2} - 3^{3/2}  \big),
\end{equation}
where $C_{1}$ and $C_{2}$ control the first- and second-invariant contributions, respectively, and $J_{m}$ controls the finite-extensibility (locking) behavior. The quantities $\bar{I}_1 - 3$ and $\bar{I}_{2}^{3/2} - 3^{3/2}$ correspond to $\xi_1$ and $\xi_2$ in Section~\ref{sec:hyperelasticity}, and the stresses are obtained from Eqs.~\ref{eq:PK2_stress_iso}--\ref{eq:PK2_to_Cauchy}. 

We consider a two-dimensional descriptor space $\bm{v} = (v_1, v_2) \in [0, 1]^{2}$, which is mapped to the modified Gent parameters $(C_1, J_m, C_2)$ through prescribed nonlinear transformations:
\begin{equation}
\begin{aligned}
C_{1} &= C_{1}^{\min} + (C_{1}^{\max} - C_{1}^{\min}) v_{1}^{0.5}, \\
C_{2} &= C_{2}^{\min} + (C_{2}^{\max} - C_{2}^{\min})v_{1}^{2} \\
J_{m} &= J_{m}^{\max} + (J_{m}^{\min} - J_{m}^{\max})\frac{1}{1 + \exp(- \kappa (v_{2} - c))}.
\end{aligned}
\end{equation}
The parameter ranges are $C_{1} \in [0.05,0.25]$, $C_{2} \in [0.01,0.05]$, and $J_{m} \in [3,20]$. We use $\kappa=15$ and $c=0.25$, creating a sigmoidal transition in $J_{m}$ centered near $v_2=0.25$. The square-root dependence of $C_1$ on $v_1$ also produces rapid changes in the first-invariant contribution near $v_1=0$. Together, these mappings create a material family in which small changes in descriptor values can produce substantially different finite-extensibility and stiffness responses. Such behavior is representative of materials in which small changes in composition or microstructure lead to pronounced changes in mechanical response. In real materials, however, constitutive responses are often coupled and may not be cleanly separable along distinct descriptor dimensions. 

\textbf{Deformation Paths:} Synthetic stress data are generated from deformation states of the form
\begin{equation}
\bm{F} = \mathrm{diag}(\lambda_{1}, \lambda_{1}^{\alpha}, (\lambda_{1} \lambda_{1}^{\alpha})^{-1}),
\end{equation}
where $\lambda_1$ denotes the first principal stretch and $\alpha \in [-0.5,1]$ parameterizes the deformation mode. The deformation is incompressible satisfying $\det(\bm{F}) = 1$. For isotropic hyperelastic materials, this provides broad coverage of the admissible isochoric invariant space. Values of $\alpha= -0.5$, $0$, and $1$ correspond to uniaxial tension, planar tension (also called pure shear), and equibiaxial tension, respectively. 

\textbf{Dataset Generation and Evaluation Task:} For each synthetic material, stress data used for training are generated along the uniaxial ($\alpha = -0.5$), pure shear ($\alpha = 0$), and equibiaxial tension ($\alpha = 1$) deformation paths. To evaluate generalization across deformation modes, the intermediate paths $\alpha = -0.25$ and $\alpha = 0.25$ are reserved for testing. For each deformation path, the admissible stretch range $[1,\lambda_1^{\max}]$ is determined by enforcing $\xi_1 < \min(0.8J_{m}$, 6.67) and $\xi_2 < 15$. The first constraint prevents sampling near the Gent locking singularity, while the latter constraint limits excessively large deformation states. Each deformation path is sampled at 128 uniformly spaced stretch values over the admissible interval.

Training data are generated by sampling descriptor values $(v_{1},v_{2})$ from the domain $[0,1]^2$ using Latin hypercube sampling (LHS). We use randomized LHS without further optimization to reflect the imperfect coverage often present in real materials datasets. The sampling procedure is repeated with different random seeds to evaluate performance across multiple stochastic training sets. The test set is generated using LHS with centered-discrepancy optimization to encourage uniform and stratified coverage of the descriptor space. This design enables model performance to be evaluated systematically across the full descriptor domain. In this work, we consider training sets containing 6, 8, 10, 12, 16, and 24 synthetic materials. Each synthetic material contains 384 stress--deformation pairs, corresponding to three training deformation modes sampled at 128 stretch values each. For each training set size, we consider 20 realizations of sampled descriptors. The test data consists of a fixed set of 64 descriptors. The synthetic benchmark evaluates generalization to unseen deformation modes, materials, and their combinations simultaneously.

\textbf{Model and Training Details:} For this benchmark, we use a monolithic pICNN with convex and conditioning hidden-layer widths of $[15,15]$, giving 872 trainable parameters. For the mixture ICNN, we use five mixture components to provide a low-dimensional set of shared energy responses while keeping the overall model size comparable to the monolithic baseline. The model uses convex hidden-layer widths of $[20,20]$ and conditioning hidden-layer widths of $[10,10]$, giving 825 trainable parameters. Here, we use the \emph{Softplus} activation function with the sharpness parameter $\beta = 1$ for both the convex and conditional branches. All models are trained for 1000 epochs with a mini-batch size of 128 using stress--deformation pairs.

\begin{figure}[!h]
    \centering
    \includegraphics[width=0.85\linewidth]{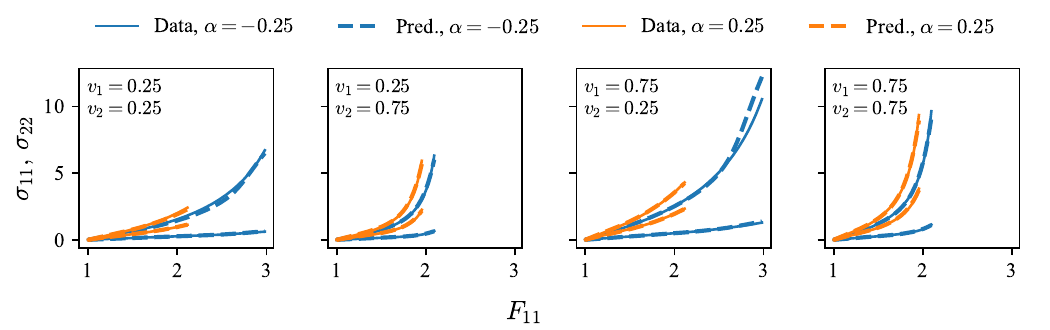}
    \caption{Representative stress predictions from the mixture model on the synthetic Gent-type benchmark. Results are evaluated at held-out descriptor values $(v_1,v_2)$ and held-out loading-mode parameters $\alpha$. Both $\sigma_{11}$ and $\sigma_{22}$ are plotted against $F_{11}$ for each value of $\alpha$. Blue traces are for $\alpha$ = -0.25 and orange traces are for $\alpha$ = 0.25.}
    \label{fig:gent_stress_stretch}
\end{figure}

\textbf{Results and Discussion:} Figure~\ref{fig:gent_stress_stretch} shows stress predictions from the mixture model on the synthetic Gent-type benchmark. The results are from a single representative run trained with 16 material descriptor pairs and evaluated at descriptor values and loading modes not included during training. The selected descriptor pairs cover low and high values of both coordinates within the descriptor domain $\bm{v} \in [0,1]^{2}$. The blue and orange curves correspond to the held-out deformation-mode parameters unseen during training, $\alpha=-0.25$ and $\alpha=0.25$, which lie between the training paths in isochoric invariant space. For each loading path, both $\sigma_{11}$ and $\sigma_{22}$ are plotted against $F_{11}$. In this example, the mixture model accurately predicts material responses across substantial differences in stiffness, locking behavior, and relative magnitudes of the two normal stress components. The curves do not extend to the same value of $F_{11}$, reflecting the fact that the admissible stretch range depends on both the synthetic material and the loading path. This is similar to experimental datasets, where measurements often cover different deformation ranges. Aggregate error metrics over multiple training-set sizes and independent runs are discussed later in this section.

\begin{figure}[!h]
    \centering
    \includegraphics[width=0.85\linewidth]{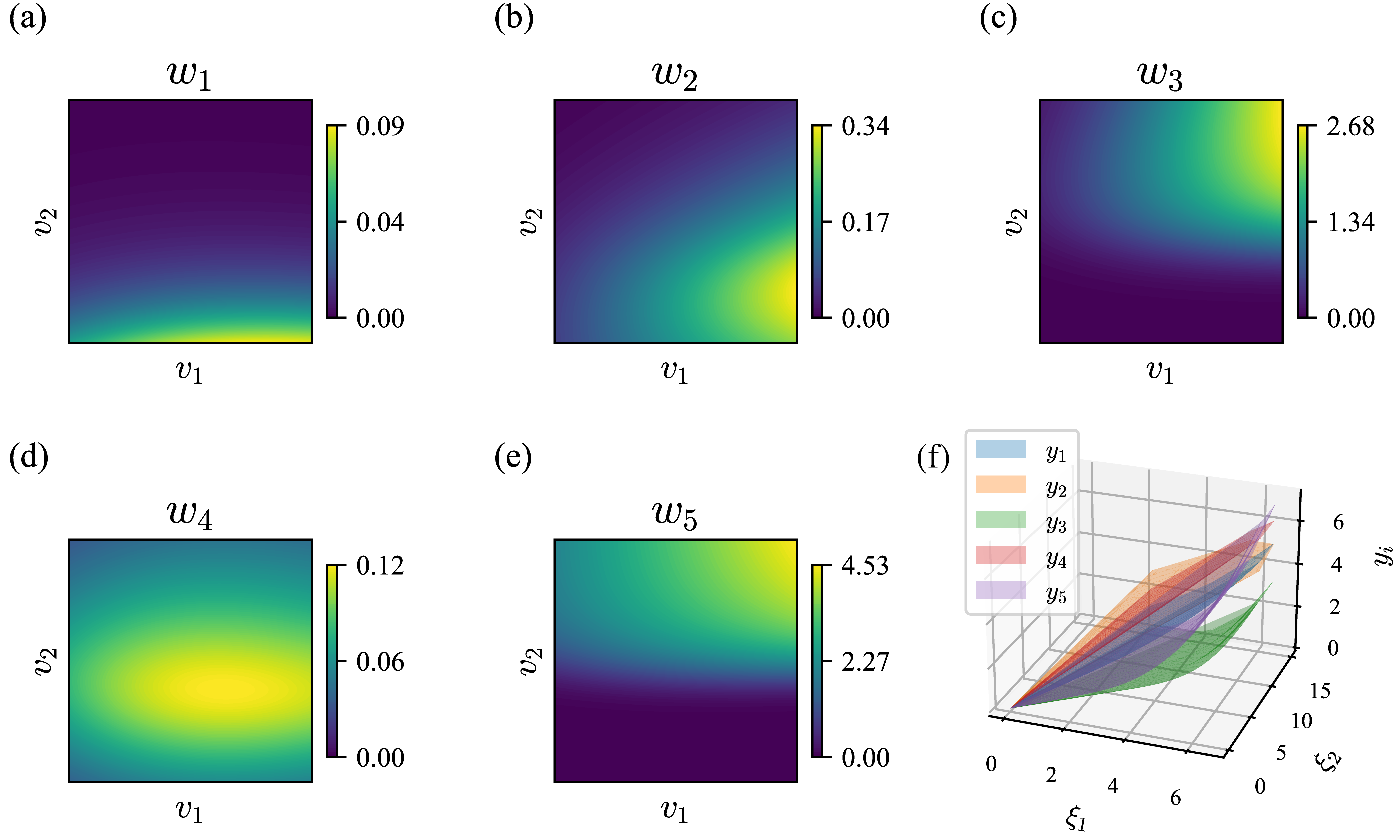}
    \caption{Learned mixture decomposition for a illustrative synthetic Gent-type benchmark run. Panels (a)--(e) show the mixture-weights $\hat{w}_{i}(v_{1},v_{2})$, and panel (f) shows the latent strain energy components $\hat{y}_{i}(\xi_{1},\xi_{2})$.}
    \label{fig:gent_learned_mixture_energy}
\end{figure}

Figure~\ref{fig:gent_learned_mixture_energy} shows the learned mixture decomposition for the same representative run discussed in the stress-prediction results above. Although this is only one possible learned decomposition, it illustrates how the mixture architecture extends to a two-dimensional descriptor space and to strain energy functions that depend on both the first and second isochoric invariant measures. The latent strain energy components in Figure~\ref{fig:gent_learned_mixture_energy}(f) exhibit different degrees of nonlinearity over the $(\xi_1,\xi_2)$ domain. Components $y_{3}$ and $y_{5}$ are non-planar energy surfaces with pronounced upturns along some loading paths. Components $y_{1}$,  $y_{2}$, and $y_{4}$ are roughly planar over much of the plotted domain. The corresponding mixture-weight functions in Figure~\ref{fig:gent_learned_mixture_energy}(a)--(e) determine where these latent energy components are used across the material descriptor space. 

In particular, the mixture weights $w_{3}$ and $w_{5}$ are close to zero when the material descriptor $v_2$ is less than 0.25. At fixed $v_{1}$, the weights increase sharply as $v_{2}$ is increased from approximately 0.25 to 0.6 and then vary more gradually thereafter. This pattern is consistent with the sigmoidal dependence of the Gent finite-extensibility parameter $J_{m}$ on $v_{2}$ in the ground-truth mapping. The sigmoid is centered at $v_{2} = 0.25$. As $v_2$ increases, the onset of locking shifts to lower stretches before saturating at $v_{2} \approx 0.6$. In this case, the learned weights provide an interpretable view of how the latent energy components capture the finite-extensibility behavior in the synthetic dataset. In addition, inspecting the learned weights can reveal underused mixture components. Here, $w_{1}$ is nearly zero across most of the descriptor domain, suggesting that fewer mixture components may be sufficient for this synthetic material family. This interpretation is supported by the mixture-size study in Appendix Figure~\ref{fig:mixture_size_study}, where the three-component mixture model achieves predictive performance comparable to that of the five-component mixture when using 16 training materials. In addition, an eight-component mixture performs similarly, indicating that modest increases in the number of mixture components do not degrade performance for the benchmark at this training-set size. 

\begin{figure}[!htb]
    \centering
    \includegraphics[width=0.8\linewidth]{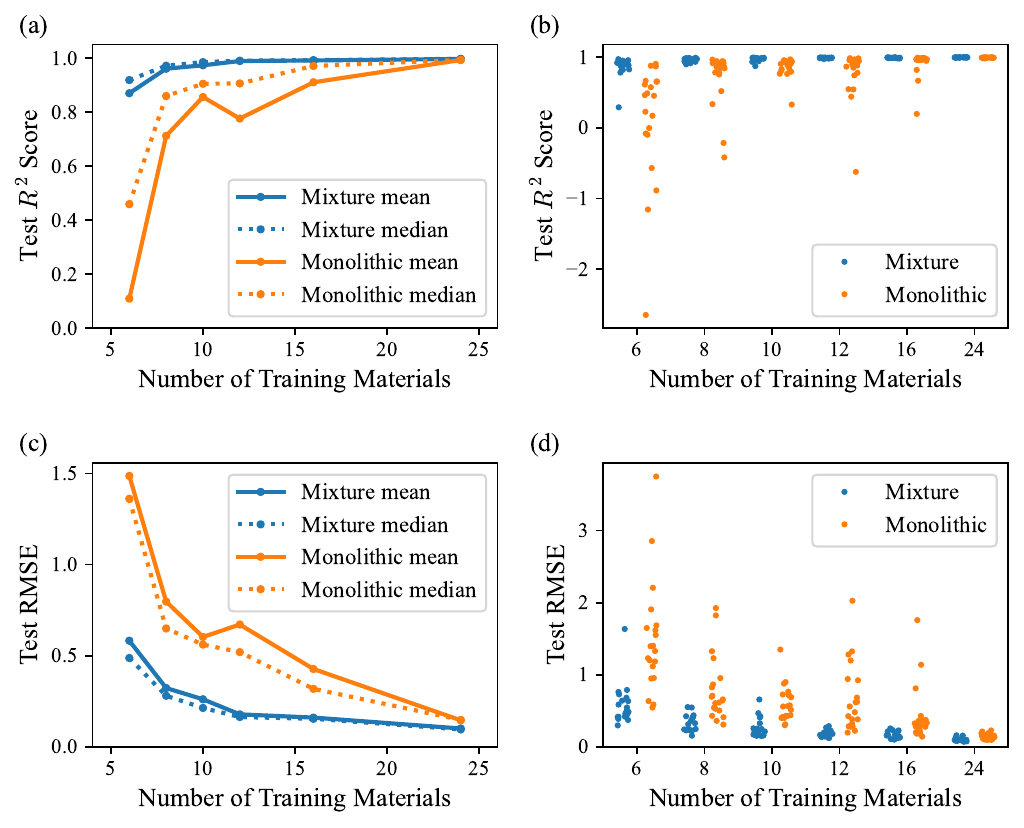}
    \caption{Test $R^{2}$ scores (a, b) and root mean squared errors (RMSE) (c, d) for joint generalization to unseen material descriptors and unseen loading modes ($\alpha=$ -0.25 and 0.25) versus the number of materials used for training. Panels (a, c) summarizes performance across 20 runs, while panels (b, d) shows the scores for individual runs. Results for each run are averaged over the final five epochs (epochs 996--1000).}
    \label{fig:R2_score_plot_996_1000_mat_mode_r2}
\end{figure}

Finally, we compare the monolithic and mixture architectures in terms of data efficiency by varying the number of materials used for training. This is important because real-world datasets are often limited by the difficulty in producing and characterizing a large number of materials. Figure~\ref{fig:R2_score_plot_996_1000_mat_mode_r2} summarizes test $R^2$ scores in panels~(a, b) and root mean squared errors (RMSE) in panels~(c, d) for joint generalization to unseen material descriptors and unseen deformation modes. For each training-set size, panel~(a, c) shows the mean and median performance score across the 20 runs, while panel~(b, d) shows the results for individual runs. Here, the mixture architecture achieves reasonable performance with as few as 12 training materials, with results improving as additional materials are included. In contrast, the monolithic architecture does not reach consistently acceptable accuracy scores until 24 seen materials. With this larger number of training materials, both architectures perform well, indicating that the advantage of the mixture architecture is reliability in small-data regimes. Critically, the monolithic architecture achieves acceptable generalization for some training-set realizations in the small-data regime, indicating that it provides a capable baseline for our study. However, its performance is less consistent and more sensitive to the particular materials descriptor included in the training set. 

Additional results in Appendix~\ref{sec:gent_appendix} consider other generalization cases and the effect of stopping training earlier. When testing on unseen material descriptors using deformation modes included in training (Appendix Figure~\ref{fig:R2_score_plot_996_1000_mat_r2}), performance follows the same overall trend as the joint-generalization case. By comparison, when evaluated on unseen deformation modes for materials seen during training (Appendix Figure~\ref{fig:R2_score_plot_996_1000_mode_r2}), both architectures achieve high test $R^{2}$ scores across all training set sizes. This indicates that, in this synthetic benchmark, the main challenge is generalization across material descriptors rather than deformation modes. As shown in Appendix Figure~\ref{fig:R2_score_plot_500_504_mat_mode_r2}, earlier stopping reduces overfitting in the monolithic architecture for smaller number of training materials, although its performance remains poor and lower than the mixture architecture. Tables~\ref{tab:appex-gent-rmse} and~\ref{tab:appex-gent-r2} summarize the RMSE and test $R^2$ scores across varying number of seen descriptors, averaged over the final five epochs of $1000$ epochs of training. 

These results should be considered in the context of the inductive bias introduced by the mixture architecture, which assumes that the material family can be captured by a small set of shared latent components. In this benchmark, this shared structure enables reliable generalization with far fewer training materials. Even for this relatively simple material descriptor mapping, the monolithic architecture requires substantially more data to achieve comparable reliability. These results demonstrate the benefits of encoding structural priors for constitutive learning from limited data. 

\FloatBarrier
\section{Conclusion}
\label{sec:conclusion}

This work addresses the challenge of predicting hyperelastic responses across a family of related materials based on composition or other descriptors. We propose a mixture of polyconvex neural potentials that learns these descriptor-dependent responses through shared latent strain energy components. Each strain energy component is represented by a neural network whose inputs are polyconvex strain invariant measures and whose architecture enforces convexity and monotonicity with respect to these inputs. A separate conditioning network maps material descriptors to non-negative mixture weights that define a weighted sum of the latent strain energy components. This construction preserves polyconvexity, which guarantees material stability and simplifies the use of the learned model in simulations. We compare the mixture architecture with the established monolithic partially input-convex neural network (pICNN) as a baseline, in which latent features derived from material descriptors are injected throughout the convex layers.

We evaluate the models on experimental data from multi-material PolyJet 3D printing and on a synthetic Gent-type benchmark. The PolyJet study uses polymer mixing ratio as a scalar material descriptor, whereas the synthetic benchmark considers a two-dimensional descriptor space and multiple deformation modes. In the PolyJet benchmark, the mixture models reliably predicted the stress responses of held-out materials across independent training runs. In contrast, the monolithic pICNN performed well for some held-out materials and training runs but did not generalize well across all cases, particularly when predicting compositions outside the training range. Critically, the performance of the mixture models was less sensitive than that of the monolithic pICNNs to network size and sharpness of the \emph{Softplus} activation functions used in the convex layers. Moreover, the synthetic Gent-type benchmark showed that the mixture model generalized more reliably under sparse sampling of the material descriptor space. The monolithic pICNN showed greater variability under sparse sampling and did not generalize consistently until the descriptor space was sampled more densely. Both architectures performed well for unseen deformation modes of materials included during training, indicating that the greater challenge was generalization across materials rather than predicting responses for a known material.

Inspection of the learned mixture decompositions shows that the latent strain energy components can specialize to capture distinct stiffness and finite-extensibility behavior. A key benefit of the mixture architecture is that the learned mixture weights can reveal how the contributions of these latent components vary across the material-descriptor space. However, these mixture decompositions are not necessarily unique and should be viewed as one possible representation of the constitutive response. The primary value of inspecting the mixture decomposition here is to support the interpretability of the trained models and not necessarily identify physical mechanisms.

The results presented here also point to several directions for extending and further evaluating the mixture-of-convex-potentials approach. Here, the approach was demonstrated only for isotropic, incompressible hyperelasticity. An important extension is to include volumetric deformation through $J=\det{\bm{F}}$ in the learned strain energy function. It would be useful to compare a single neural network that jointly represents the distortional and volumetric responses with a modular construction in which these contributions are represented by separate networks and combined additively. In addition, the number of mixture components is specified in advance. Future work could explore sparsity-promoting gating mechanisms or other adaptive methods that allow the effective number of components to be determined from the data. In addition, reducing model complexity through network sparsification could be investigated to assess whether it improves the generalization performance of both architectures. Beyond these model extensions, the present benchmarks are limited to relatively small material families with one- and two-dimensional descriptor spaces. Future work should test the approach on larger families with higher-dimensional descriptor spaces and a wider range of constitutive responses to determine whether the mixture structure improves generalization or limits model flexibility. Such extensions could also consider how the material-descriptor space influences other properties, such as strength, and whether certain compositions can be manufactured. This could motivate the development of constraint-aware parametric material models that restrict material selection to feasible regions of the descriptor space.

More broadly, mixtures of convex potentials could be extended to inelastic constitutive models, where they could be used to represent a strain energy potential governing the recoverable response, a dissipation potential, or both. The same idea could also be applied to other parametric modeling problems where convex potentials provide useful physical or mathematical structure. Overall, these results show that the shared mixture structure can improve generalization across material families, particularly when only limited data are available.

\section*{Acknowledgments}
This work was supported by the Laboratory Directed Research and Development program (project 233095) at Sandia National Laboratories, a multimission laboratory managed and operated by National Technology and Engineering Solutions of Sandia LLC, a wholly owned subsidiary of Honeywell International Inc. for the U.S. Department of Energy’s National Nuclear Security Administration under contract DE-NA0003525.
 
This paper describes objective technical results and analysis. Any subjective views or opinions that might be expressed in the paper do not necessarily represent the views of the U.S. Department of Energy or the United States Government.

\section*{Declaration of generative AI and AI-assisted technologies use}
During the preparation of this work, the authors used generative AI and AI-assisted
tools, including ChatGPT for language editing of the manuscript text. Additionally, these tools were utilized to assist in generating code for plotting our results. The authors thoroughly reviewed, verified, and edited all outputs to ensure factual
and technical accuracy, and take full responsibility for the final content of the publication.

\FloatBarrier
\bibliography{references}  

\FloatBarrier
\appendix
\FloatBarrier
\section{Appendix}
\label{sec:appendix}

\setcounter{figure}{0}
\renewcommand{\thefigure}{A\arabic{figure}}

\setcounter{table}{0}
\renewcommand{\thetable}{A\arabic{table}}

\FloatBarrier
\subsection{Experimental Multi-material 3D Printing Benchmark}

\begin{table}[!h]
\centering
\caption{Training and test RMSE values and test $R^2$ scores averaged across multiple runs with three random seeds and six cross-validation folds. Interpolation and extrapolation metrics are averaged separately over four interior and two edge test folds, respectively. Here S refers
to the small network and L refers to the large network as described
in Section~\ref{sec:experimental_multimaterial_3d_printing}}
\label{tab:mm3dp_metrics}
\small
\begin{tabular}{lllllllll}
\toprule
Model      & Config. & \begin{tabular}[c]{@{}l@{}}Train \\ RMSE\end{tabular} & \begin{tabular}[c]{@{}l@{}}Test \\ RMSE\end{tabular} & \begin{tabular}[c]{@{}l@{}}Interp. \\ RMSE\end{tabular} & \begin{tabular}[c]{@{}l@{}}Extrap. \\ RMSE\end{tabular} & \begin{tabular}[c]{@{}l@{}}Test \\ $R^2$\end{tabular} & \begin{tabular}[c]{@{}l@{}}Interp. \\ $R^2$\end{tabular} & \begin{tabular}[c]{@{}l@{}}Extrap. \\ $R^2$\end{tabular} \\ \midrule
Mixture    & S-0.5   & 0.069                                                 & 0.225                                                & 0.142                                                   & 0.390                                                   & 0.95                                                  & 0.98                                                     & 0.89                                                     \\
           & S-1.0   & 0.065                                                 & 0.201                                                & 0.119                                                   & 0.364                                                   & 0.96                                                  & 0.99                                                     & 0.90                                                     \\
           & S-2.0   & 0.061                                                 & 0.185                                                & 0.118                                                   & 0.320                                                   & 0.97                                                  & 0.99                                                     & 0.93                                                     \\ \cmidrule{2-9} 
           & L-0.5   & 0.069                                                 & 0.254                                                & 0.151                                                   & 0.460                                                   & 0.94                                                  & 0.98                                                     & 0.85                                                     \\
           & L-1.0   & 0.065                                                 & 0.259                                                & 0.180                                                   & 0.418                                                   & 0.94                                                  & 0.97                                                     & 0.88                                                     \\
           & L-2.0   & 0.061                                                 & 0.243                                                & 0.157                                                   & 0.415                                                   & 0.94                                                  & 0.97                                                     & 0.88                                                     \\ \midrule
Monolithic & S-0.5   & 0.075                                                 & 0.411                                                & 0.306                                                   & 0.619                                                   & 0.83                                                  & 0.89                                                     & 0.71                                                     \\
           & S-1.0   & 0.064                                                 & 0.592                                                & 0.450                                                   & 0.875                                                   & 0.63                                                  & 0.72                                                     & 0.47                                                     \\
           & S-2.0   & 0.060                                                 & 0.523                                                & 0.418                                                   & 0.733                                                   & 0.72                                                  & 0.78                                                     & 0.61                                                     \\ \cmidrule{2-9} 
           & L-0.5   & 0.075                                                 & 0.491                                                & 0.342                                                   & 0.787                                                   & 0.75                                                  & 0.85                                                     & 0.55                                                     \\
           & L-1.0   & 0.063                                                 & 0.598                                                & 0.443                                                   & 0.910                                                   & 0.60                                                  & 0.69                                                     & 0.43                                                     \\
           & L-2.0   & 0.059                                                 & 0.722                                                & 0.607                                                   & 0.951                                                   & 0.38                                                  & 0.38                                                     & 0.37                                                     \\ \bottomrule
\end{tabular}
\end{table}

\begin{figure}[h]
    \centering
    \includegraphics[width=0.8\linewidth]{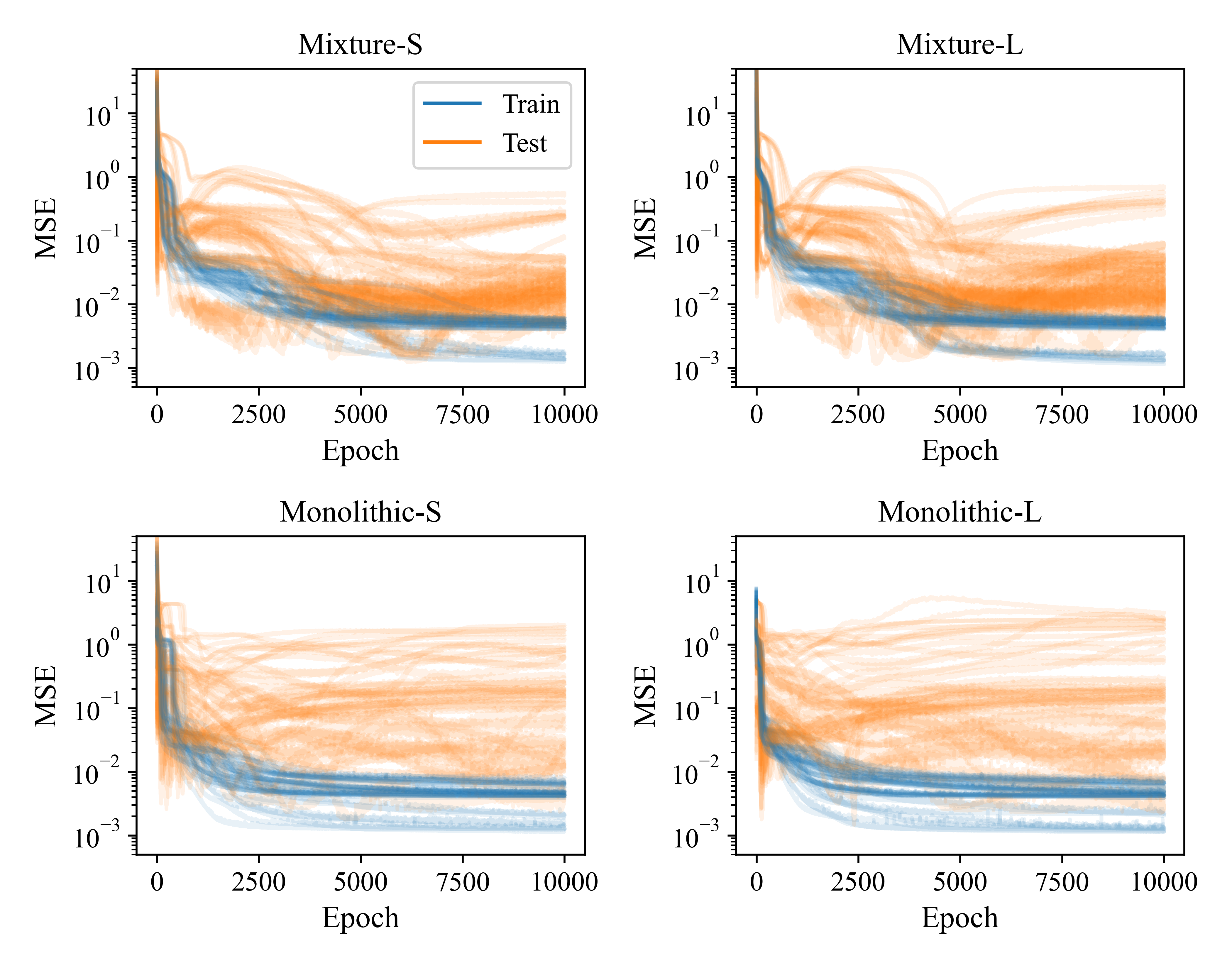}
    \caption{Train and Test loss curves for experimental multi-material 3D printing benchmark for mixture and monolithic networks.}
    \label{fig:mm3dp_training_history}
\end{figure}

\FloatBarrier
\subsection{Synthetic Gent-Type Benchmark}

\begin{figure}[!h]
\centering
\includegraphics[width=0.8\linewidth]{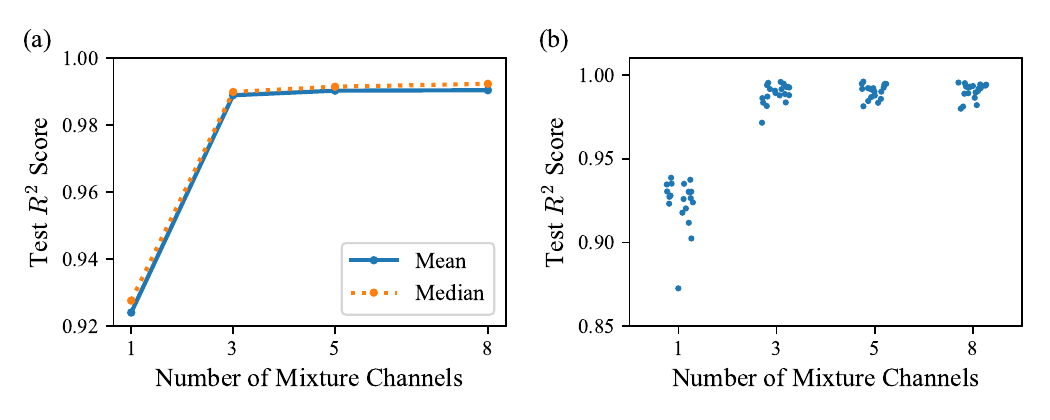}
\caption{Mixture-size study for joint generalization to unseen material descriptors and unseen loading modes ($\alpha=$ -0.25 and 0.25) on the Gent-type benchmark using 16 training material descriptors. The ICNN and conditioning network have hidden layer dimensions $[20,20]$ and $[10,10]$, respectively, while the number of mixture components is varied. (a) Mean and median test ($R^2$) score across random seeds versus the number of mixture components. (b) Test ($R^2$) scores for the individual random seeds.}
\label{fig:mixture_size_study}
\end{figure}

\begin{figure}[!h]
    \centering
    \includegraphics[width=0.8\linewidth]{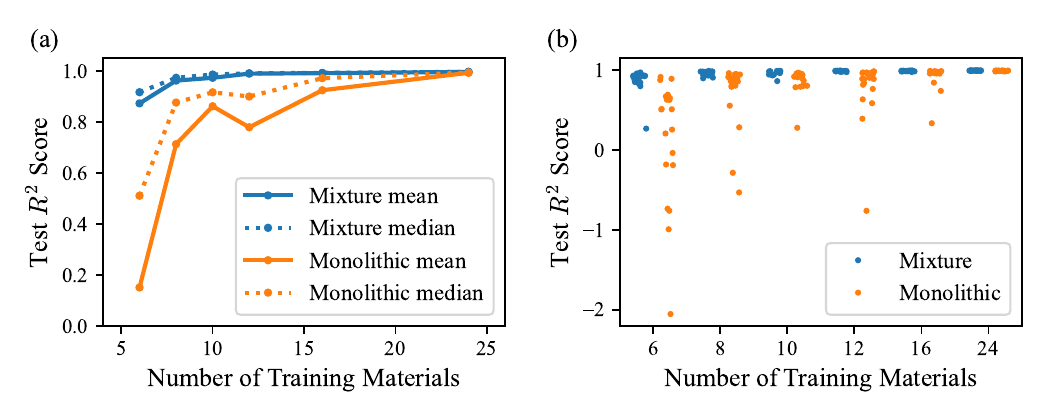}
    \caption{Test $R^{2}$ scores for generalization to unseen material descriptors under seen loading modes ($\alpha=$ -0.5, 0.0, and 1.0), versus number of materials used for training. Panel (a) summarizes performance across 20 runs, while panel (b) shows the scores for individual runs. Results for each run are averaged over the final five epochs (epochs 996--1000).}
    \label{fig:R2_score_plot_996_1000_mat_r2}
\end{figure}

\begin{figure}[!h]
    \centering
    \includegraphics[width=0.8\linewidth]{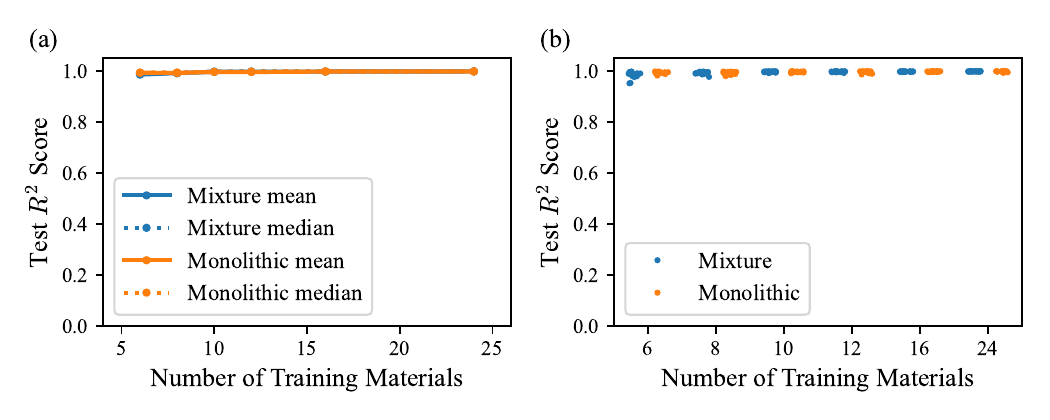}
    \caption{Test $R^{2}$ scores for generalization to unseen loading modes ($\alpha=$ -0.25 and 0.25) for material descriptors seen during training. Results are plotted for number of training materials. Panel (a) summarizes performance across 20 runs, while panel (b) shows the scores for individual runs. Results for each run are averaged over the final five epochs (epochs 996--1000).}
    \label{fig:R2_score_plot_996_1000_mode_r2}
\end{figure}

\label{sec:gent_appendix}
\begin{figure}[!h]
    \centering
    \includegraphics[width=0.8\linewidth]{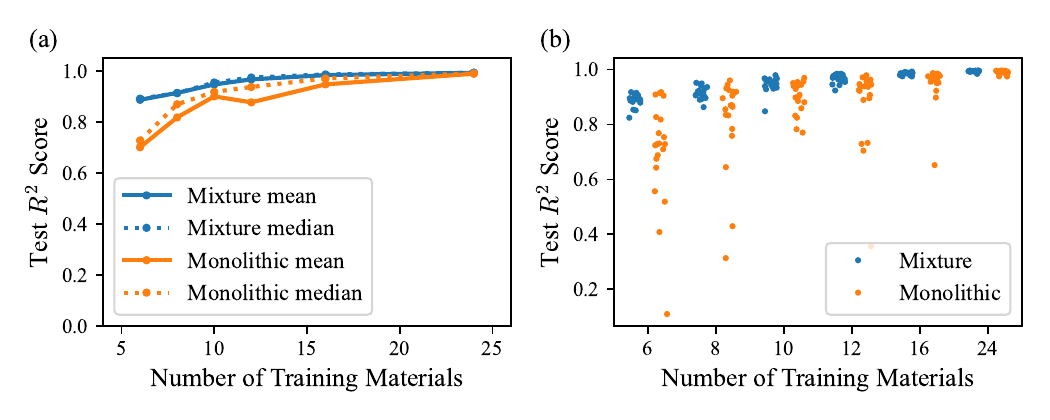}
    \caption{Test $R^{2}$ score for the early-stopping case by averaging results over epoch 500--504. Results are shown for joint generalization to unseen material descriptors and unseen loading modes ($\alpha=$ -0.25 and 0.25) versus the number of materials used for training. Panel (a) summarizes performance across 20 runs, while panel (b) shows the scores for individual runs.}
    \label{fig:R2_score_plot_500_504_mat_mode_r2}
\end{figure}

\begin{table}[h]
\centering
\caption{Training RMSE and test RMSE values for the mixture and monolithic networks for different numbers of materials used during training. Test RMSE is reported for joint generalization to unseen material descriptors and loading modes (Mat. + Mode), unseen material descriptors under seen loading modes (Mat.), and unseen loading modes for seen material descriptors (Mode). The reported values are averaged over the final five epochs (epochs 996--1000) and across 20 independent training-set realizations.}
\small
\label{tab:appex-gent-rmse}

\begin{tabular}{llllll}
\toprule
Model      & $N_{\mathrm{mat}}$ & Train RMSE & Test RMSE (Mat. + Mode) & Test RMSE (Mat.) & Test RMSE (Mode) \\ \midrule
Mixture    & 6                  & 0.1663     & 0.5823                  & 0.5403           & 0.1759           \\
           & 8                  & 0.1281     & 0.3223                  & 0.2973           & 0.1382           \\
           & 10                 & 0.0738     & 0.2605                  & 0.2431           & 0.0781           \\
           & 12                 & 0.0749     & 0.1779                  & 0.1607           & 0.0798           \\
           & 16                 & 0.0627     & 0.1601                  & 0.1447           & 0.0674           \\
           & 24                 & 0.0543     & 0.1002                  & 0.0911           & 0.058            \\ \midrule
Monolithic & 6                  & 0.1067     & 1.4871                  & 1.3327           & 0.1137           \\
           & 8                  & 0.1097     & 0.7971                  & 0.7166           & 0.1209           \\
           & 10                 & 0.074      & 0.6029                  & 0.539            & 0.0814           \\
           & 12                 & 0.0795     & 0.671                   & 0.5995           & 0.0869           \\
           & 16                 & 0.0586     & 0.4269                  & 0.3661           & 0.0624           \\
           & 24                 & 0.0558     & 0.1451                  & 0.1291           & 0.0596           \\ \bottomrule
\end{tabular}
\end{table}

\begin{table}[h]
\centering
\caption{Test $R^{2}$ values for the mixture and monolithic networks for different numbers of materials used during training. Test RMSE is reported for joint generalization to unseen material descriptors and loading modes (Mat. + Mode), unseen material descriptors under seen loading modes (Mat.), and unseen loading modes for seen material descriptors (Mode). The reported values are averaged over the final five epochs (epochs 996--1000) and across 20 independent training-set realizations.}
\small
\label{tab:appex-gent-r2}
\begin{tabular}{lllll}
\toprule
Model      & $N_{\mathrm{mat}}$ & Test $R^{2}$ (Mat. + Mode) & Test $R^{2}$ (Mat.) & Test $R^{2}$ (Mode) \\ \toprule
Mixture    & 6                  & 0.869                      & 0.873               & 0.986               \\
           & 8                  & 0.96                       & 0.962               & 0.992               \\
           & 10                 & 0.972                      & 0.973               & 0.997               \\
           & 12                 & 0.988                      & 0.99                & 0.997               \\
           & 16                 & 0.99                       & 0.992               & 0.998               \\
           & 24                 & 0.996                      & 0.997               & 0.999               \\ \hline
Monolithic & 6                  & 0.108                      & 0.152               & 0.993               \\
           & 8                  & 0.711                      & 0.713               & 0.992               \\
           & 10                 & 0.855                      & 0.862               & 0.996               \\
           & 12                 & 0.775                      & 0.78                & 0.996               \\
           & 16                 & 0.909                      & 0.925               & 0.998               \\
           & 24                 & 0.992                      & 0.993               & 0.998               \\ \bottomrule
\end{tabular}
\end{table}

\end{document}